\documentclass{desyproc}
\usepackage{paralist}
\usepackage{subfig}
\usepackage{array}

\begin{document}
\title{Status report and first results of the microwave LSW experiment at CERN}

\author{{\slshape M. Betz$^1$, F. Caspers$^1$, M. Gasior$^1$, M. Thumm$^{2}$}\\[1ex]
$^1$European Organization for Nuclear Research (CERN), Geneve, Switzerland\\
$^2$Karlsruhe Institute of Technology (KIT), Germany}

\contribID{Betz\_Michael}

\desyproc{tba}
\acronym{Patras 2012} 
\doi  
\linespread{1.0}
\maketitle

\begin{abstract}
To detect or exclude the existence of hidden sector photons or axion like particles, 
a table-top ``microwaves shining through the wall'' experiment has been set up at CERN.
An overview of the experimental layout is given, the technical challenges involved are reviewed and the
measurement procedure including data-evaluation and its results to date are shown.
  
\end{abstract}

\section{Introduction}
Axion like particles (ALPs) and Hidden Sector Photons (HSP) belong to the family of Weakly Interacting Sub-eV Particles (WISPs), which
can be probed by ``Light Shining through the Wall'' (LSW) experiments, exploiting their weak coupling to ordinary photons.
This allows an indirect proof of the otherwise hidden
particles without relying on any cosmological assumptions.
Previous LSW experiments have been carried out with laser light at DESY (ALPS), CERN (OSQAR)
and Fermilab (GammeV).

\section{Engineering aspects}
\begin{figure}[tbh]
\begin{center}
\subfloat[Block-diagram]{
\includegraphics[width=0.55\textwidth]{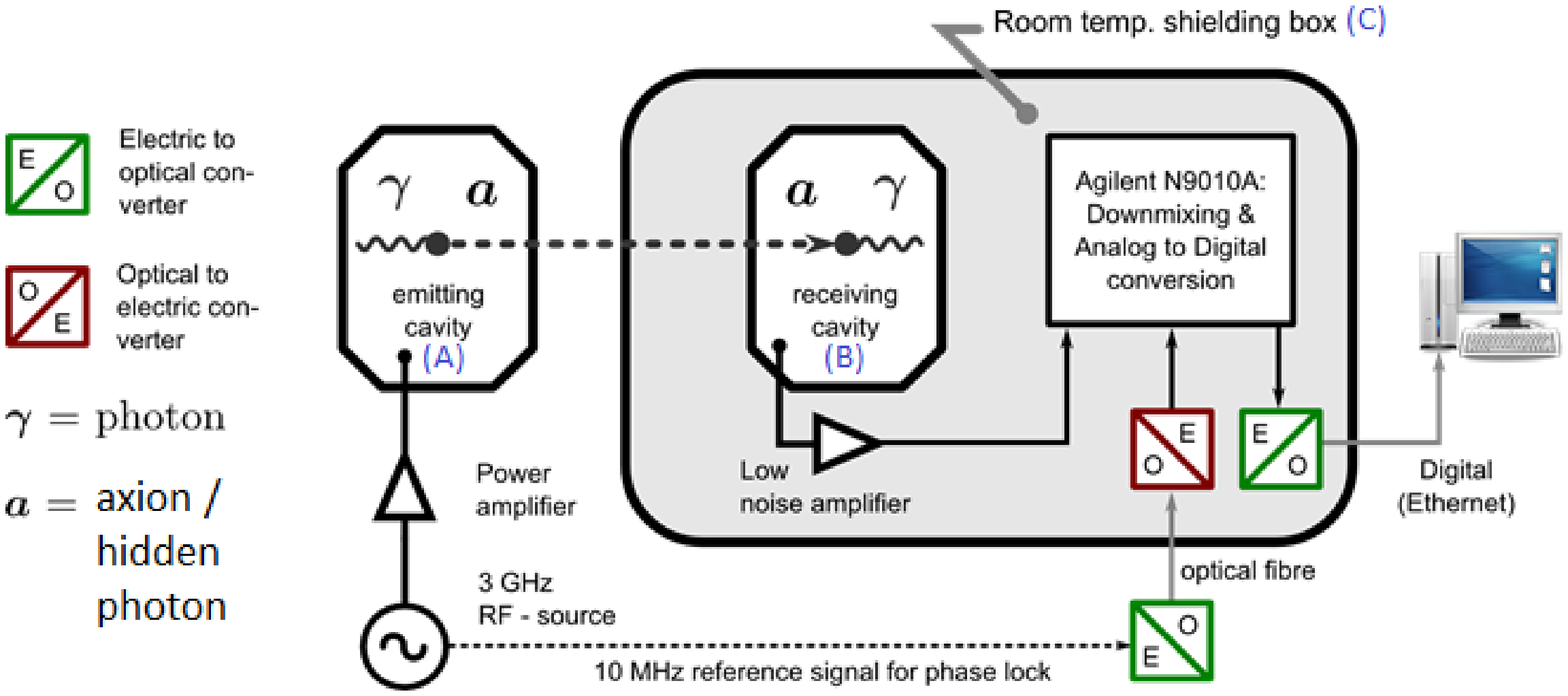}
\label{fig:ovrBlock}
}
\subfloat[Actual setup]{
\includegraphics[width=0.4\textwidth]{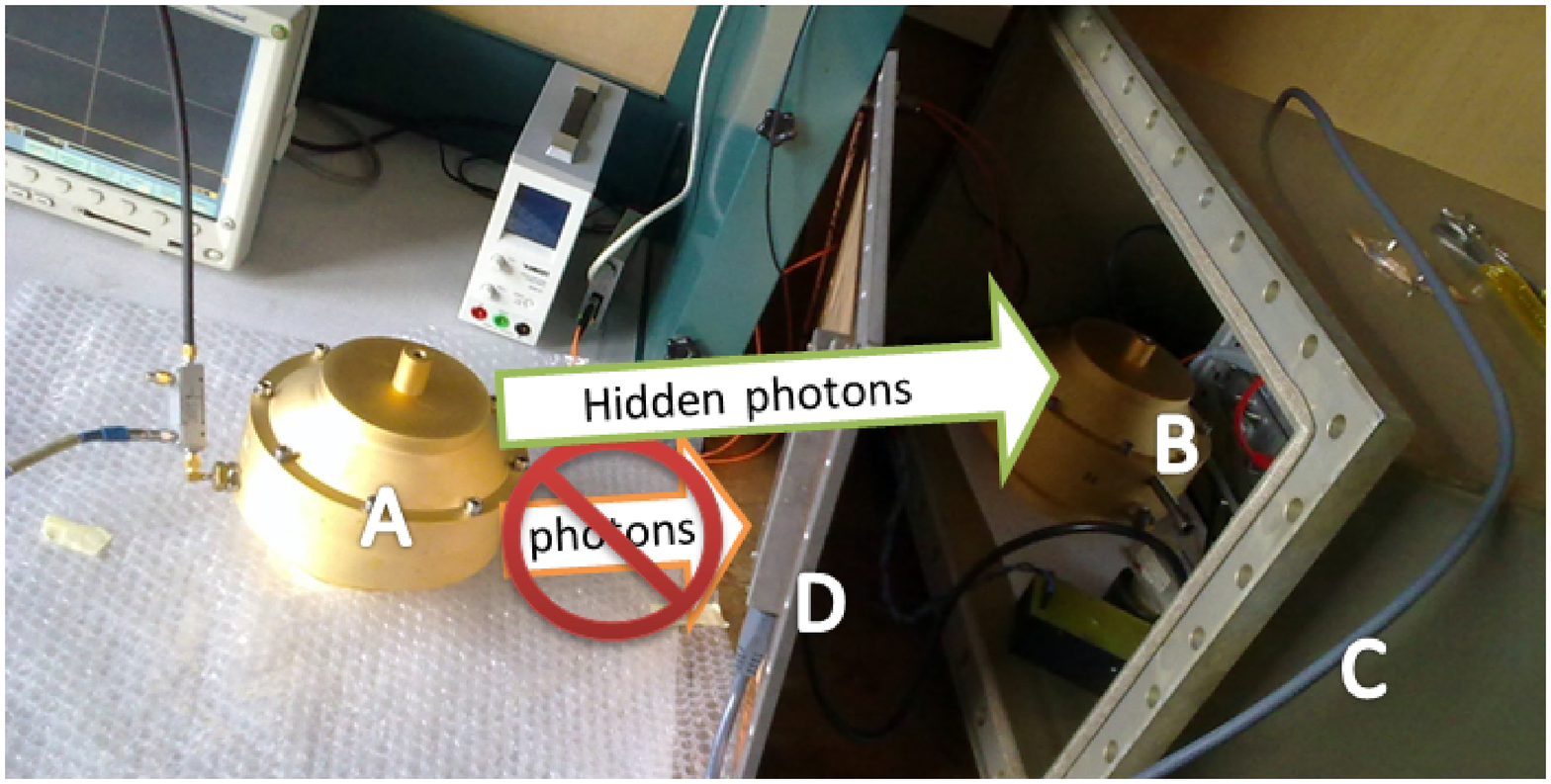}
\label{fig:ovrPhoto}
}
\end{center}
\caption{Overview of the light shining through the wall experiment in the microwave range}
\label{fig:theExp}
\vspace{-.5cm}
\end{figure}

The concept of an optical LSW experiment \cite{src:moiIPAC} has been adapted to microwaves as shown in Fig.~\ref{fig:theExp}.
Two identical microwave cavities with a diameter of 140 mm each are positioned in close vicinity to each other.
One serves as emitting cavity and is excited on its resonant frequency with $P_{em}$ = 50 W of RF power at $f_{sys}$ = 2.9565 GHz by
an external microwave source.
Theory predicts that microwave photons from the emitting cavity can convert to HSPs by kinetic mixing or to ALPs by the
Primakoff effect, penetrate the cavity walls and convert back to photons in the detection cavity \cite{src:JaCaRi}.
Considering current exclusion limits, the probability for this process to happen is $< 10^{-25}$.
If a small excitation of the detection cavity over the thermal noise level can be observed and if any direct electromagnetic
(EM) crosstalk can be excluded, this could indicate the existence of WISPs.

The detecting cavity is placed in a EM shielding enclosure, together with a Low Noise Amplifier (LNA) and 
a Vector Signal Analyzer (VSA).
The EM shielding is a critical aspect of the experiment because
it is not possible to distinguish between a signal originating from WISPs and direct coupling
between the cavities by EM leakage. The nominal electric field strength inside the emitting cavity
is 186 kV/m. The smallest detectable signal in a 12 h run corresponds to an electric field strength of 24 nV/m in
the detection cavity. 
So the shielding must attenuate the EM signal by at least 258 dB.
The cavity walls provide 110 dB each, which was measured by a calibrated near field probe. 
The limiting factor is the contact resistance between the two half-shells.
The additional EM shielding box provides 90 dB of
attenuation. The overall screening attenuation thus is $> 310$ dB, which is impossible to measure in its combined form, 
but which has been determined shell by shell.
All analog and digital signals are transmitted by fibre optic transceivers, to ensure interference can not couple
through the screening layer of coaxial cables. The VSA is remote controlled by a fibre optic ethernet link.
To provide AC power within the shielding box, a lowpass feed-trough filter has been constructed for 50 Hz mains,
attenuating interference in the MHz region and above by $>$ 90 dB.

The resonant frequency of both cavities must stay close to the system frequency $f_{sys}$ during the
measurement run.
We define half the 3~dB bandwidth of the cavities (65~kHz) as the limit for the maximum amount of
allowable drift.
The emitting cavity dissipates 50~W of power by forced air cooling. It heats up and expands considerably. During a warm up period 
the drift of $\approx$~1~MHz is manually compensated by a tuning screw. After $\approx$~1 h, thermal equilibrium is reached, the cavity is
stable and no further tuning adjustments are necessary. We monitor the reflected power from its
coupling port continuously by a directional coupler and a detector diode,
 as long it is below a certain threshold, the cavity is considered to be on tune.
The resonant frequency of the detecting cavity is monitored by observing a bump in its spectral
noise power density.
A low noise amplifier (LNA), based on a High Electron Mobility Transistor (HEMT) is directly connected to the detecting cavity,
it provides G = 44.7~dB gain at $T_{LNA}$ = 32.4~K
 noise temperature. The noise temperature of the detection cavity is frequency dependent, its maximum is
 at resonance and equals the physical temperature of $T_{CAV}$ = 290~K. 
 This can be observed as a significant peak over the noise floor of the
 amplifier, allowing us to determine the resonant frequency. The overall system noise temperature of the complete receiving chain, 
 determined by Friis formula is $T_{SYS}$ = 32.5~K.

Since there is no energy loss associated with the WISP conversion process, the regenerated photons
in the detecting cavity appear with the same energy as the photons in the emitting cavity. Thus,
the signal which is coupled out from the detecting cavity has the same frequency and bandwidth as the one which is generated on
the emitting side. For this reason, data processing is a matter of detecting a sinusoidal signal of known frequency in the
white background noise.
The VSA records complex samples, representing a bandpass limited signal with up to 25 MHz span, centered at $\approx f_{sys}$.
The recorded data is evaluated by a python script, estimating the noise power spectrum.
After applying a Hanning window to reduce the effect of spectral leakage 
(if the narrowband signal falls between two bins and no window is used, its amplitude can be reduced by up to 36 \% \cite{src:fftPaper}),
one Fast Fourier Transform (FFT) is calculated over the entire time trace,
which can be up to several hundred millions of samples long.
 The ``FFTW'' software library is used to implement the FFT efficiently.
\begin{wrapfigure}{r}{0.4\textwidth}
\centerline{\includegraphics[width=0.4\textwidth]{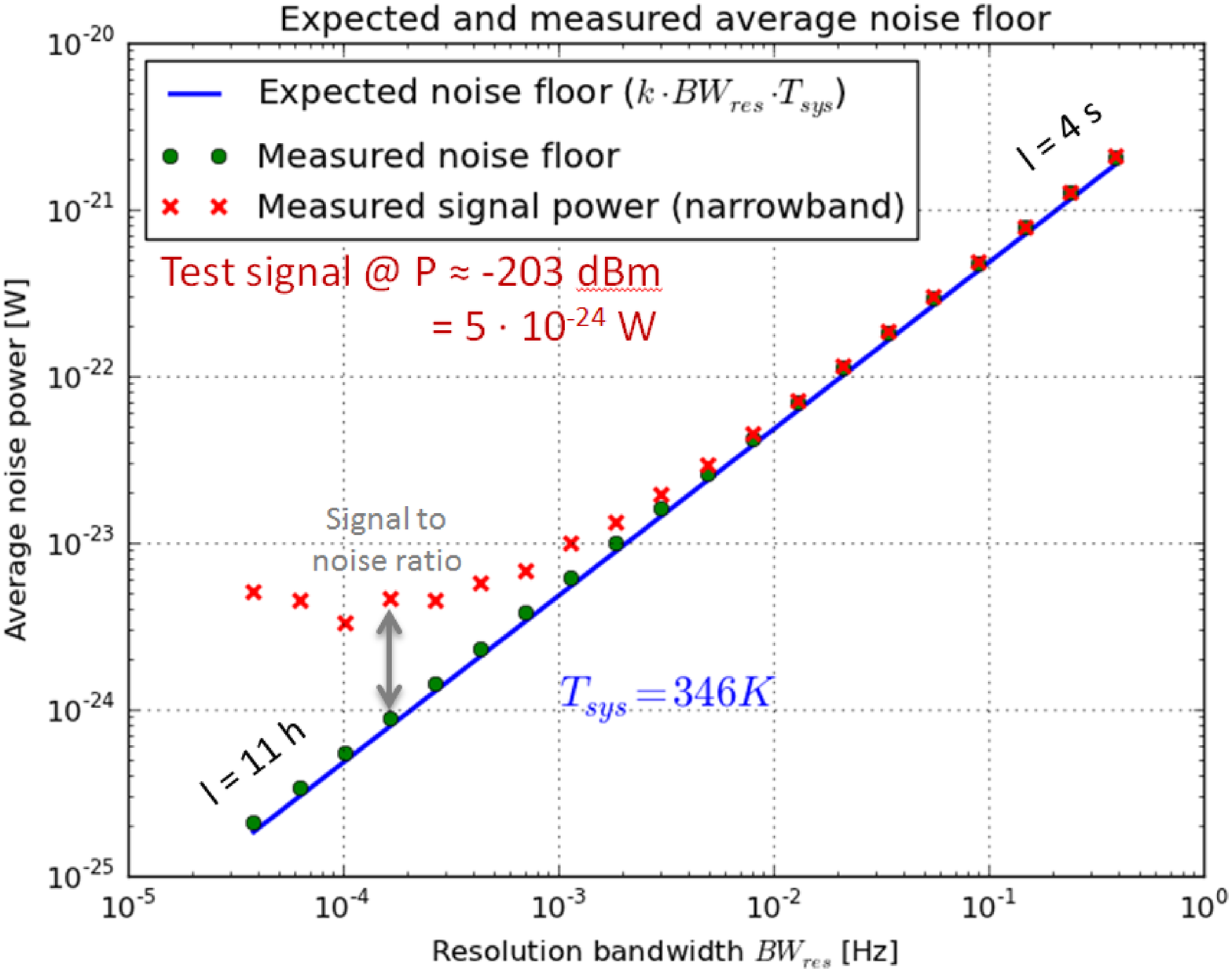}}
\caption{Decrease of the average noise floor with longer time traces and thus narrower resolution bandwidth.}
\label{fig:sigNoise}
\end{wrapfigure}
Considering pure white noise as an input signal to the calculation, $\mathrm{BW}_{\mathrm{res}} = \mathrm{BW}_{\mathrm{wnd}} / \tau$
describes the 3 dB bandwidth of one spectral bin. It depends on the length of the recorded time trace ($\tau$) and the
applied window function. The Hanning window broadens the frequency response by a factor of $\mathrm{BW}_{\mathrm{wnd}} = 1.5$.
The average available noise power $P_n = k_B \mathrm{BW}_{\mathrm{res}} \mathrm{T}_{\mathrm{sys}}$
decreases proportionally with the length of the time trace $\tau$.
If a signal is observed, it will appear as a single line in the spectrum with constant power. The signal
 to noise ration should thus improve linearly with $\tau$.
For this to hold true, all oscillators must be synchronized and stay
stable in frequency relative to each other within a tolerance given by $\mathrm{BW}_{\mathrm{res}}$.
Synchronization is achieved by a 10 MHz signal, which all instruments use as frequency reference. 
Long-term frequency drifts in the order of few mHz can not be avoided during the $\tau > 12$~h measurement time.
However, with synchronization, they will affect all oscillators in the same way and cancel out in
the measurement result.
This way, each bin of the synchronized FFT effectively corresponds to a lock-in amplifier, tuned to the bin's center frequency.
To proof the setup is stable over 12 h, a test run has been conducted. For the evaluation, its time trace has been truncated
to different lengths.
In Fig~\ref{fig:sigNoise}, the expected linear decrease of the average noise floor can be observed. A narrowband
signal has been introduced by provoking electromagnetic leakage. Once the signal 
emerges from the noise floor, it stays relatively constant in power. 
The signal peak never exceeded the width of two bins in the spectrum, which is the minimum dictated by the Hanning window, showing that the
frequency-locking with a relative accuracy in the $\mu$Hz range has been achieved between all the involved oscillators.
\vspace{-4mm}
\section{Measurement runs}
A 11.5 h measurement run for HSPs has been carried out in March 2012. No narrowband signal was detected and the data was interpreted
as an exclusion result. Details can be found in \cite{src:moiIPAC}. We improved over the most
sensitive exclusion limits from cosmic microwave background measurements, in the energy range of $10~\mu$eV. 

For axion measurements, both cavities need to be placed in a strong magnetic field. Thus modification
of the setup was necessary and a smaller secondary shielding box has
been constructed from non magnetic stainless steel. It contains the detecting cavity, the LNA and a
analog optical link with 4.5 GHz bandwidth. The optical fibre allows signal transmission from the secondary to the primary
shielding box, containing the optical receiver and the VSA. The cavities were tuned to the TM$_{010}$ mode,
at 1.76 GHz, coupling well to axions \cite{src:JaCaRi}.
The hardware transfer function from the cavity port, through all cables, the LNA, and the optical link has been carefully determined
by a network analyzer to be 60.0 dB at 2.95 GHz and 58.7 dB at 1.76 GHz, allowing
to normalize the measured noise spectra.

\begin{figure}[tbh]
\begin{center}
\subfloat[Power spectrum]{
\label{fig:exclSignal}
\includegraphics[width=0.33\textwidth]{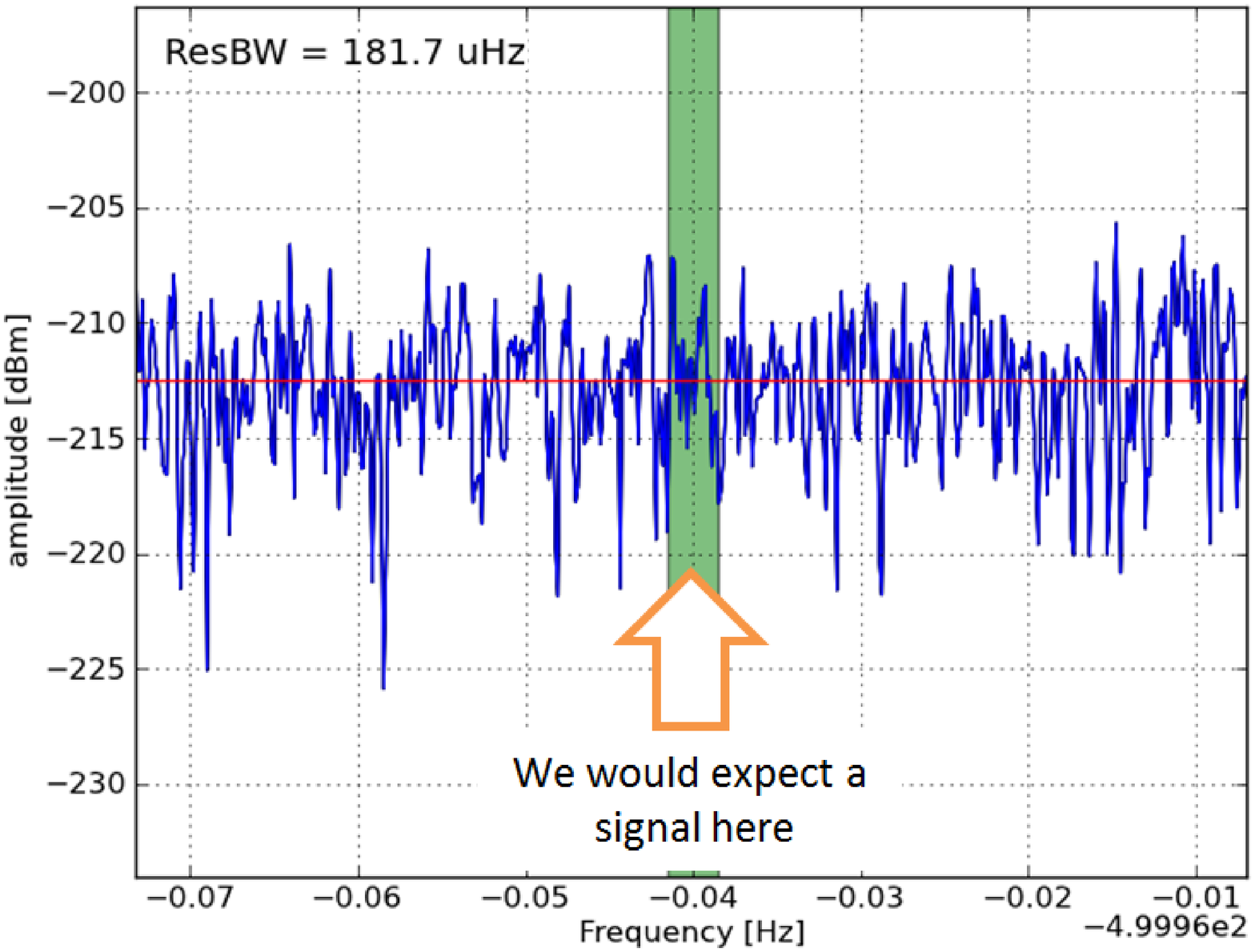}
}
\subfloat[Derived exclusion limit]{
\label{fig:exclDetail}
\includegraphics[width=0.302\textwidth]{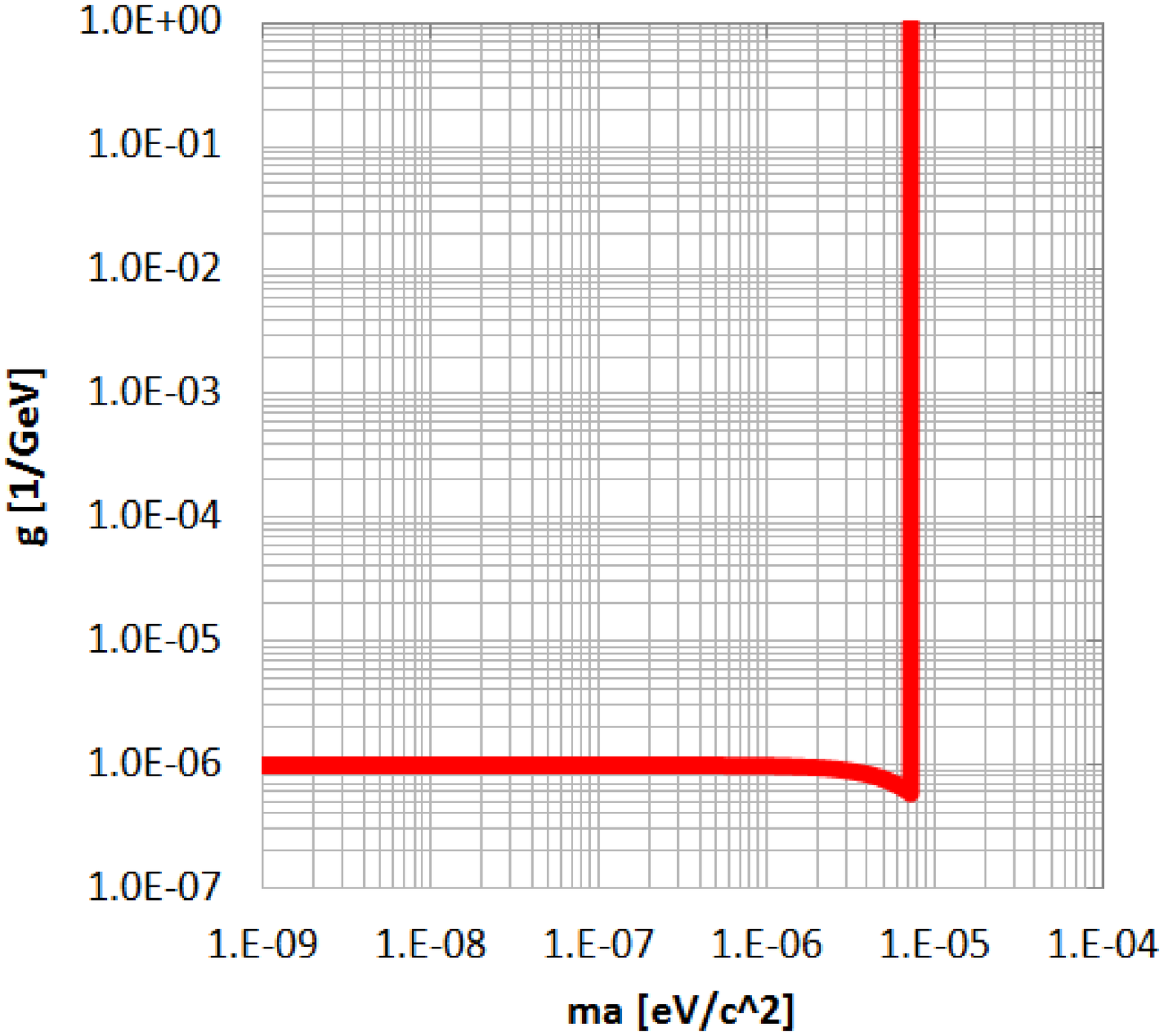}
}
\subfloat[Compared to other experiments]{
\label{fig:exclOverview}
\includegraphics[width=0.33\textwidth]{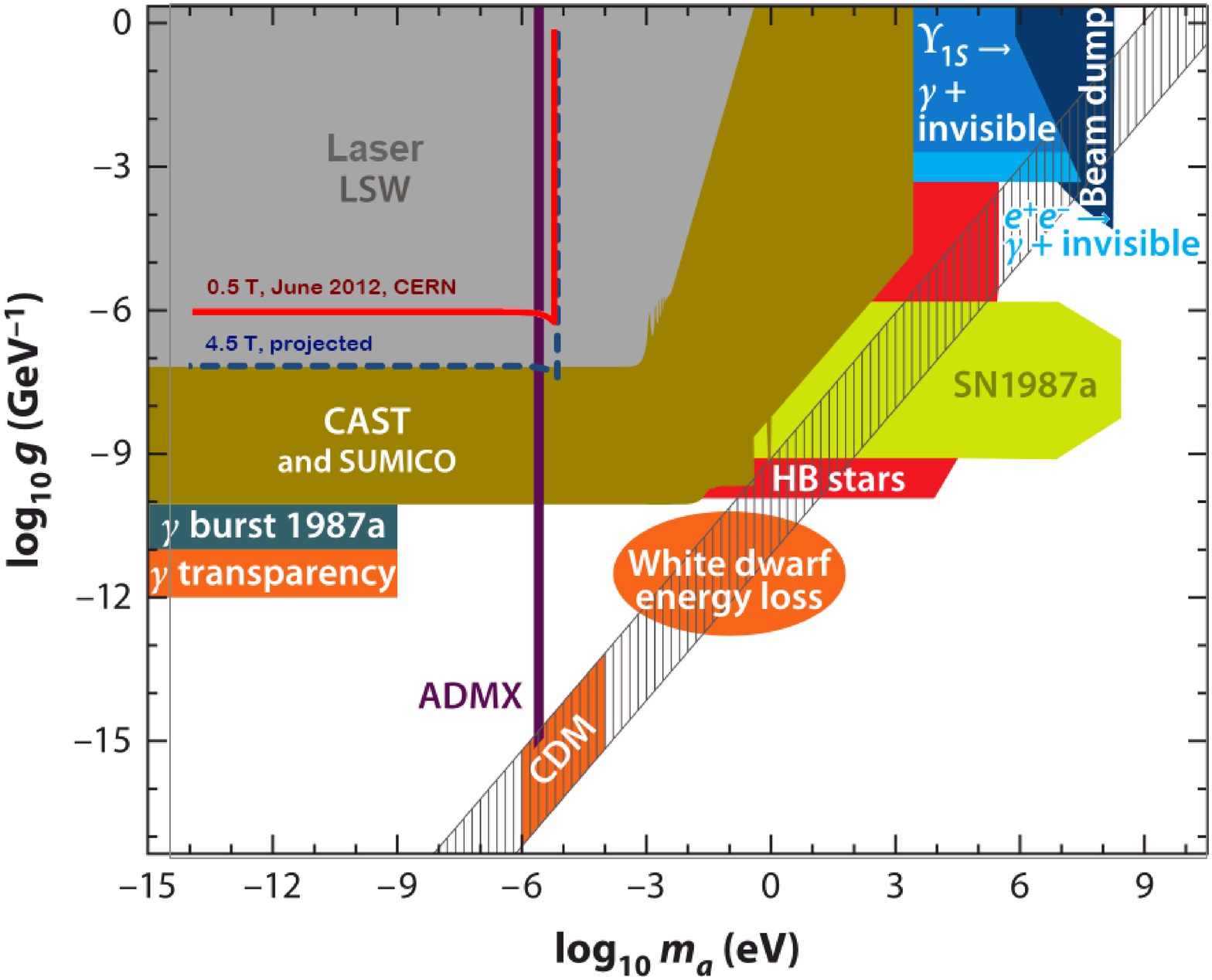}
}
\end{center}
\vspace{-4mm}
\caption{Exclusion limits for axions derived from the second 4 h measurement run (see text)}
\label{fig:exclAxion}
\end{figure}

In June 2012 a magnet was available during a one week timeslot.
The normal conducting dipole magnet is regularly used for material testing at CERN and has a aperture of $\approx$~100~x~50~x~30~cm.
It provides an average field strength of 0.51 T, which has been confirmed by a calibrated hall probe.
We were able to record 3 measurement runs of 4 h, 4 h and 6 h length. 
The first run showed EM leakage. A weak spot could be narrowed down with a near-field probe to the gasket between lid and
flange of the enclosure and mitigated by copper mesh.
No signal and only noise was observed in the second and third run (see Fig.~\ref{fig:exclSignal}). The data
of the second run was translated to an exclusion limit for axions and is shown in 
Fig.~\ref{fig:exclDetail} and Fig.~\ref{fig:exclOverview} as the red trace.
In the third run, $\approx$ 7 mm$^3$ of Cr$_2$O$_3$ crystals have been placed
on the bottom of each cavity. One may speculate if Cr$_2$O$_3$ interacts with axions because of its 
magneto-electric properties \cite{src:fritzPatras}. However, despite a 2\% reduction in Q factor, no surprising effects were observed. 
\vspace{-7mm}
\vspace{-5mm}
\section{Conclusion and Outlook}
Although current exclusion limits could not be challenged, we 
gained experience on our measurement setup and successfully operated it within a magnet for the first time. This allows to utilize a
stronger, superconducting magnet in a next step.
We are planning to work together with an external partner, where a superconducting magnet with 
a wide enough aperture for our measurement setup is available. The projected sensitivity with this
magnet (4.5 T) and the current setup in a 8 h run is shown in Fig.~\ref{fig:exclOverview} as the blue trace.
\\
{\small
The authors would like to thank R.~Jones, E.~Jensen and the BE department management for encouragement and support.
Thanks to the organizers of the Patras Workshop for a very enjoyable and inspiring conference.
Supported by the Wolfgang-Gentner-Programme of the Bundesministerium f\"ur Bildung und Forschung
(BMBF).
}


\begin{footnotesize}

\end{footnotesize}



\begin{thebibliography}{9}   
\bibitem{src:moiIPAC}
M. Betz, F. Caspers, ``\textit{A microwave paraphoton and axion detection experiment with 300 dB electromagnetic shielding at 3 GHz}'',
proc. IPAC 2012, New Orleans, CERN-ATS-2012-089


\bibitem{src:JaCaRi}
J.~Jaeckel, A.~Ringwald, \textit{``A Cavity Experiment to Search for Hidden Sector Photons''}, Physics B659 2008


\bibitem{src:fftPaper}
G. Heinzel et al., 
``\textit{Spectrum and spectral density estimation by the Discrete Fourier transform (DFT), including a comprehensive list of window
functions and some new flat-top windows}'', Max-Planck-Institut f\"ur Gravitationsphysik 02/2002

\bibitem{src:fritzPatras}
F. Caspers, ``\textit{Engineering aspects of microwave axion experiments}'', proc. of Patras 2010, Z\"urich
\end{thebibliography}
\end{document}